\documentclass{aastex}
\usepackage{graphicx}
\shorttitle{More galaxies in the Local Volume imaged in $H{\alpha}$}
\shortauthors{Karachentsev \& Kaisin}

\begin{document}

\title{More galaxies in the Local Volume imaged in $H{\alpha}$}
\author{Igor D.\ Karachentsev
and Serafim S.\ Kaisin}
\affil{Special Astrophysical Observatory, Russian Academy
	  of Sciences, N.\ Arkhyz, KChR, 369167, Russia}
\email{ikar@sao.ru}
\begin{abstract}
  We have carried out an $H\alpha$ flux measurement for 52 nearby galaxies
as part of a general $H\alpha$ imaging survey for the Local Volume sample
of galaxies within 10 Mpc. Most of the objects are probable members of
the groups around Maffei~2/IC 342, NGC 672/IC 1727, NGC 784, and the Orion
galaxy. The measured $H\alpha$ fluxes corrected for extinction are used to
derive the galaxy star formation rate (SFR). We briefly discuss some basic
scaling relations between SFR, hydrogen mass and absolute magnitude of the
Local Volume galaxies. The total SFR density in the local $(z = 0)$
universe is estimated to be $(0.019\pm0.003)M_{\odot}$ yr$^{-1}Mpc^{-3}$.

\end{abstract}

\section{Introduction}

Systematic measurements of $H\alpha$ fluxes in nearby galaxies
within a fixed distance is one of the major techniques for
determining the star formation rate (SFR) in the local universe.
The presence of dwarf galaxies with extremely
low luminosities in the Local Volume which are usually
invisible at large distances, provides a
unique opportunity for researching the SFR of a galaxy depending
on its luminosity, structure, gas mass, and density of its
environment in the broadest possible range of these
characteristics. Thus it is essential that the observational
program contains objects of all types and sizes in order to avoid
the observational selection, distorting the interpretation of
results.

Kraan-Korteweg \& Tammann (1979) proposed to view an exemplary
sample of 179 nearest galaxies with distances within 10 Mpc
(the so-called LV sample). Later, Karachentsev (1994)
replenished it with galaxies discovered from new redshift
surveys. The updated sample constituted  226 galaxies.
Later, Karachentseva \& Karachentsev (1998,2000) and Karachentseva
et al. (1999) undertook the all-sky search for nearby galaxies
using plates of the POSS-II and ESO-SERC survey. A large number
of new nearby dwarf galaxies of low surface brightness were found.
A limiting magnitude of the survey was $B \simeq 17^m$ providing
an essential completeness of the LV sample up to
absolute magnitude $M_B \simeq -12.5^m$ within a distance of 8 Mpc.
Their radial velocities  were measured by Huchtmeier et al. (2000,
2001) during a subsequent HI-survey. Distances to many of the LV
 galaxies were first measured with high accuracy on the Hubble
Space Telescope (HST) using the tip  red giant branch (TRGB) method.
The results of these international efforts were summarized in the
Catalog of Neighboring Galaxies (CNG) (Karachentsev et al., 2004),
consisting of 451 galaxies with distance estimates within $D<10$
Mpc.

Different surveys of large areas of the sky in the optical and radio bands:
SDSS (Abazajian et al. 2009), 2dF (Colless et al. 2001), 6dF (Jones et al. 2004),
HIPASS (Zwaan et al. 2003), and ALFALFA (Giovanelli et al. 2005) have led to
an  essential increase of the Local Volume sample. The second version of
the CNG (Karachentsev et al. 2011, in preparation) contains $N\simeq790$ galaxies, i.e.
more than four times  that of the original list by
Kraan-Korteweg \& Tammann (1979). Such a representative sample,
being registered in the $H\alpha$ emission line, provides a detailed
picture of star formation in the Local Volume during the recent
time interval of $\sim$10 Myr.

Measurements of $H\alpha$ fluxes in nearby galaxies were performed
by many authors. Their interest was normally fixed on objects
of a certain type, for example, on spiral or irregular galaxies. A
synopsis of publications, where the number of measured $H\alpha$
fluxes for the Local Volume galaxies was not less than 10, is
presented in Table 1. The first column of the table contains a
link to the paper, the second contains the number of LV galaxies imaged
in $H\alpha$, and the third column reflects the nature of the sample
(the morphological type of galaxies or their affiliation to a
fixed group). The last row of the table summarizes the number of LV
galaxies, which were observed in our program (Karachentsev et al.
2005; Kaisin \& Karachentsev 2006, 2008; Kaisin et al. 2007;
Karachentsev \& Kaisin 2007),
including the results of this paper. Unlike all previous
observational programs, our $H\alpha$ survey of LV galaxies does
not imply any selection of objects by morphological type, or their
affiliation to a group. This has led to some unexpected results,
in particular, to the detection of circumnuclear emission in
isolated E/S0 galaxies (Moiseev et al. 2010).

As follows from the data above, so far there are 692 measurements
of the $H\alpha$ flux available for 435 LV galaxies, i.e. a lot of
galaxies were observed independently by different authors, which enables
an estimate of the external error of the flux measurement. More
than  half of the $H\alpha$ data were obtained within the
framework of two programs: Kennicutt et al. (2008) and our survey.

The degree of completeness of the Local Volume galaxy collection
currently available remains quite ambiguous. New sky surveys
reveal new nearby galaxies of both low and high surface
brightness. Refinement of individual distances to galaxies leads
to their inclusion or exclusion as LV members. Among
$\sim790$ galaxies that are currently listed in the LV, there are
objects with absolute magnitudes $M_B$ ranging from $-22^m$ to
$-4^m$, i.e. their luminosities vary by more than seven orders of
magnitude.  Figure 1 presents the distribution of 572 galaxies
situated within 8 Mpc according to
their B-band absolute magnitude. The shaded areas on the
left and right panels indicate numbers of galaxies that have been
observed in $H\alpha$ and $HI$, respectively. The median absolute
magnitude of the LV galaxies is $-14^m$. Measurements of the
$H\alpha$ flux are carried out for 355 galaxies or 62\% of the sample.
As one can see, the current $H\alpha$ survey is almost complete on the
bright half of the luminosity function, i.e. up to $M_B\simeq-15^m$.
For a comparison note that the completeness of the LV galaxy survey
in the neutral hydrogen line HI is much higher, reaching 88\%.
A comparison of these panels indicates the need to speed up the survey
of nearby galaxies  in the $H\alpha$ line to have a sufficiently
complete picture of the SFR diversity in them.

\section{Observations and image processing}

Below we present a survey of 52 nearby galaxies, most of which
were observed in the $H\alpha$ line for the first time. Some of
these galaxies are likely members of the closest neighboring group
Maffei2/IC342, poorly studied due to its location in the region of
strong absorption in the Milky Way; the other part is a mixture of
field galaxies and members of other small groups, in
particular, around NGC~672, NGC~784, and in the Orion region.

CCD images of galaxies in the $H\alpha$-line and in the continuum
were obtained during observing runs from 2005 to 2009. An average
seeing was $1.8\arcsec$. All the observations were performed in
the Special Astrophysical Observatory of the Russian Academy of
Sciences (SAO RAS) with the BTA 6-m telescope equipped with the
SCORPIO focal reducer (Afanasiev et al.\ 2005). A CCD chip of
2048$\times$2048 pixels provides a total field of view of about
6.1$\arcmin$ with a scale of 0.18$\arcsec$/pixel. The images in
$H\alpha$+[NII] were obtained via observing the galaxies
though a narrow-band interference filter $H\alpha
(\Delta\lambda=$75\AA) with an effective wavelength
$\lambda$=6555\AA.  In order to remove the stellar continuum
contribution to these images, we also observed the same ields
with two medium-band filters situated on both sides from $H\alpha$:
SED607 with $\lambda$=6063\AA, $\Delta\lambda$=167\AA, and
SED707 with $\lambda$=7063\AA,\, $\Delta\lambda$=207\AA.
Typical exposure times were $2\times600$s in $H\alpha$ and
$2\times300s$ in the continuum. Since the range
of radial velocities in our sample is small, we
used  the same $H\alpha$ filter for all the observed
objects. Our data reduction followed the standard practice and was
performed within the MIDAS package. For all the data, bias was
subtracted and the images were flat-fielded by twilight flats.
Cosmic particles were removed and the sky background was
subtracted. The next operation was to  spatially align all
the images for a given object. Then the images in the continuum
were normalized to $H\alpha$ images using 5--15 field stars and
were subtracted. $H\alpha$ fluxes were obtained for the
continuum-subtracted images, using spectrophotometric standard
stars from Oke (1990) observed during the same nights as the
objects. The investigation of measurement errors, brought in by
the continuum subtraction, flat-fielding, and scatter in the
zeropoints, has shown that they have typical values within  15\%.
We did not correct $H\alpha$ fluxes for the contribution of the
[NII] lines, because it is likely to be small for the majority of
low-luminosity galaxies in our sample. For instance, according
to Equation (1B) in Kennicutt et al. (2008), a typical galaxy in our
sample with the median absolute magnitude $M_B= -15^m$
has a [NII]/$H\alpha$ ratio of 1/20 which is lower
than the accuracy of our measurements.

\section{$H\alpha$ fluxes and SFRs}

The measured integrated flux of a galaxy in the $H\alpha$+[NII]
lines, calibrated according to Oke's spectrophotometric standards
(and noted as $F(H\alpha))$,
was expressed in units of (erg$\times$ cm$^{-2}\times$ sec$^{-1})$.
 In each case, we tried to take into account not only the sum
of emission knots of the galaxy but also its diffuse emission
background insofar as it was not distorted by the background
subtraction errors, significant in large apertures. Due to the
width of the filter  used, the measured galaxy fluxes are also
containing emission in the neighboring line doublet [NII].
However, according to Kennicutt et al. (2008), relative
contribution of the doublet for the majority of galaxies is small,
especially for dwarf galaxies. The measured $F(H\alpha)$ flux is
then corrected for light absorption in the Milky Way $A_B$(MW)
using a technique by Schlegel et al. (1998), and for internal
extinction in the galaxy itself $A_B$(int), defined as

$$A_B({\rm int})=[1.6+2.8(\log V_m-2.2)]\times \log(a/b),\eqno(1)$$
if $V_m>42.7$ km s$^{-1}$, and $A$(int)=0 otherwise. This ratio
incorporated the known fact (Verheijen 2001) that internal
extinction depends not only on the inclination of the galaxy,
expressed in terms of its apparent axial ratio $a/b$, but also on
its luminosity, an indicator of which according to Tully \& Fisher
(1977) is the amplitude of its rotation velocity $V_m$.  Here, the
quantities $a/b$, $V_m$, and $A_B$(int) are taken from Tables 1 and 4
of the CNG. Absorption in the $H\alpha$ line was
adopted as proportional to the absorption in the $B$-band:

$$A(H\alpha)=0.538[A_B({\rm MW})+A_B({\rm int})].\eqno(2) $$

Following Gallagher et al. (1984), we calculated the integrated
star formation rate in the galaxy as

$$ SFR(M_{\odot}/year)=1.27\times 10^9\times F_c(H\alpha)\times D^2, \eqno(3) $$

where $D$ is the distance to the galaxy, expressed in Mpc. The
validity of the linear transition (3) from the flux $F_c(H\alpha)$
to the SFR has recently been a subject of critical reviews.
Pflamm-Altenburg et al. (2007) and Pflamm-Altenburg \& Kroupa
(2009) exhaustively argued that the canonical relation (3)
underestimates the SFR value in dwarf galaxies, and the
weaker the luminosity of the galaxy is, the stronger the difference
is. In dwarf systems with  absolute magnitude $M_B\sim -10, -12^m$, an
underestimation of the SFR value may reach one to two orders of
magnitude. To be pricise, in what follows we conserve the SFR
estimates made under the canonical relation (3). We divided the
galaxies we observed into two categories: 19 galaxies belonging to
a nearby association around Maffei2/IC342 and 33 general field
galaxies that include both isolated galaxies and members of
multiple systems. The galaxies Maffei1, Maffei2, and their
companions are obscured from us by dust clouds, which create a
strong and heterogeneous absorption, reaching $A_B\sim5^m-7^m$.
This circumstance considerably hampers the measurements of
$F(H\alpha)$ fluxes and increases the errors. Moreover, the
magnitude of absorption $A_B$ itself is determined at low galactic
latitudes $\mid b\mid<5^{\circ}$ with a high uncertainty.

The images of 19 galaxies from the Maffei2/IC342 group and 33
general field galaxies are represented as a mosaic in Figures 2 and
3, respectively. The left and right images of each galaxy
correspond to the sum and difference of the images that have been
exposed in the $H\alpha$ and in the continuum. The image scales
are shown by the horizontal bars equal to 1 arcmin, and the
north and east orientation is indicated in the corner by arrows.

The main characteristics of the observed galaxies are listed in
Tables 2 and 3, the structures of which are identical. The table
columns contain: (1) --- galaxy name, (2) --- equatorial coordinates
for epoch J2000.0, (3) --- morphological type from de
Vaucouleurs digital classification, (4) --- radial velocity relative
to the Local Group centroid, (5) --- distance to the galaxy
presented in the CNG catalog including new data from
Tully et al. (2006) and Karachentsev et al. (2006), (6) --- absolute
magnitude in the $B$ band from the CNG, corrected for internal and
external absorption, (7) --- logarithm of the hydrogen mass of the
galaxy, $\log(M_{HI}/M_{\odot})= \log F_{HI}+2\log D+5.37$, where
$F_{HI}$ is the flux in the HI line (in $Jy$ km $ s^{-1}$), and $D$
in Mpc, (8,9) --- the observed and corrected flux in the $H\alpha$
line, and (10) --- integrated star formation rate in the galaxy
calculated from the canonical relation (3).

\section{Comments on individual objects in the Maffei group}

The structure and kinematics of the binary association of galaxies
around the major spirals Maffei2 and IC342 as subgroup centers
were considered by Karachentsev et al. (2003a). Most of the
galaxies presented in Table 2 are physical companions of either
Maffei2 or IC342, judging from their radial velocities.

{\em KKH5, KKH6.} Both dIrr galaxies are peripheral members of
the association, that have probably not yet reached the virialized
region around Maffei2. Their distances are measured with an
accuracy of $\sim10$\% from the  TRGB
detected on the images derived with the Hubble Space Telescope (Karachentsev
et al. 2003b, 2006). Both galaxies show emission knots, which are more
diffuse in the case of KKH6.

{\em Cas1 = KK19.} This dIrr galaxy is located in the zone of
strong absorption $(A_B=4.4^m)$, which remains yet inaccessible for
determining the distance via the TRGB. Its distance as a companion
of IC 342 is adopted at 3.3 Mpc. The $H\alpha$ image reveals many
compact HII regions some of which are circular-shaped.

{\em KKH11, KKH12, MB1 = KK21.}
All the three irregular galaxies are companions of  Maffei2. Each
galaxy demonstrates the presence of emission knots and filaments
with the values of the total $H\alpha$ flux close to one another.

{\em Maffei1.}
Maffei1 is an elliptical galaxy, the distance to which is estimated by
Fingerhut et al. (2003) from the central 
dispersion of radial velocities. Its
angular sizes, corrected for absorption $A_B=5.05^m$, extend
beyond our image frame, which created problems with background
subtraction. Apart from the residues of overexposed star images,
there are no visible traces of emission regions on the body of the
galaxy. In the table we indicate the upper limit of its possible
$H\alpha$ flux.

{\em Maffei2.}
This barred spiral (Sbc) galaxy also extends beyond our image frame.
Based on its apparent $K_s$ magnitude ($5.22^m$) from the
Two Micron All Sky Survey (2MASS)  and
from the width of the HI line (305 km s$^{-1}$), we estimated its
distance as 3.1 Mpc using the infrared Tully--Fisher relation
$M_K=-9.35\lg(W_{50})+0.75$. We obtained the total $H\alpha$ flux
of Maffei2 assuming that the $H\alpha$ profile of the galaxy
reproduces its brightness profile in the $K$--band whereas about
50\% of the luminosity appeared to be outside of our frame.

{\em Dwingeloo2.} Dwingeloo2 is  an irregular galaxy in the region of strong
absorption ($A_B=5.12^m$). A rather bright star is projected on
it. The image of this star is likely screening a significant part
of the galaxy's $H\alpha$ flux.

{\em MB3 = KK22.} For this dIrr galaxy, we estimated only the upper
limit of its $H\alpha$ flux.

{\em Dwingeloo1 = KK23 = Cas2.}
This is a spiral SBbc-type galaxy, prone to strong absorption
$(A_B=6.34^m)$. For it we adopted the mean distance of the group
3.0 Mpc, although an individual assessment of the distance from
the infrared Tully-Fisher relation gives (at $K_s$ = 8.82 and
 $W$ = 187 km s$^{-1}$) a distance of 4.7 Mpc. Approximately 20\% of the
luminosity of the galaxy is beyond the range of our image.

{\em KK35.}
This object looks like an isolated spot of low surface brightness
at the distance of 16 arcmin from the center of IC 342. It can
be a dIrr galaxy in the process of merging with the giant spiral
IC342 or a stellar association on the outer spiral pattern of the
galaxy. Its distance, 3.16 Mpc, determined via the TRGB, is
consistent within errors with the distance of IC342 itself, $3.28$
Mpc, defined via the Cepheids. As the $H\alpha$ image shows, KK35
is in a state of very active star formation.

{\em UGCA86.}
This is an Sm/Im galaxy with a bright association of blue stars
(VIIZw9) on the SE side. Its distance, 2.96 Mpc, measured via
the TRGB (Karachentsev et al. 2006), confirms the affiliation of
UGCA86 to the IC 342 companions. Compact emission knots and filaments
are scattered throughout the galaxy disk, but more than half
of the integrated $H\alpha$ flux comes from a powerful site of
star formation, VIIZw9.

{\em CamA = KK41, CamB = KK44.}
Both are dIrr galaxies of low surface brightness that show the presence of
a blue stellar population in the images obtained with the WFPC2 on
the HST (Karachentsev et al. 2003a). Both galaxies show small
compact knots and a weak diffuse emission component in the
$H\alpha$ line, which indicates a sluggish process of star
formation in these systems.

{\em NGC1569.}
Along with M82 and several Markarian galaxies, NGC1569 belongs to
the objects of the Local Volume with the most active star
formation per luminosity unit of the galaxy. An abundance of
arc-like  emission filaments in the periphery makes
the galaxy resemble a crab. Due to significant absorption
($A_B=3.02^m$) the distance to NGC1569 has long remained
uncertain. Recently Grocholski et al. (2008) determined its
distance as 3.36 Mpc using the TRGB method.

{\em UGCA92.}
This dIrr galaxy is located in the sky in the vicinity of the
previous galaxy. Radial velocities and distances from the observer
of UGCA92 and NGC1569 are also close. These galaxies can form a
bound pair on the outskirts of the group around IC342, reminiscent
of the famous pair of NGC 147 and NGC 185, which are dwarf galaxies in the
neighborhood of M31. In the UGCA92 body there are groups of bright
emission knots, and two arcs interlocking in the northern part of
the galaxy.

{\em NGC1560, UGCA105.}
These are two galaxies of late Sd and Sm types, oriented  at different
angles to the line of sight. Their $H\alpha$ fluxes are also
approximately equal. Their characteristic features are ring-shaped
HII regions, apart from which there are many compact emission
knots.

\section{Comments on individual objects in the field}

{\em NGC404.}
This is the nearest  isolated lenticular galaxy of moderate
luminosity. In its central part small dust clouds are visible. Del
Rio et al. (2004) found an extended HI shell around NGC404, and
ultraviolet observations with  GALEX have identified a
ring-shaped structure of young stars (Thilker et al. 2010). On our
$H\alpha$ image one can see fairly bright emission in the
circumnuclear region of the galaxy, as well as separate emission
knots, scattered in the periphery. The integrated $H\alpha$ flux
of the galaxy, shown in Table 3, does not have any contribution of
distant emission knots in the HI region shell. Recently, Williams et al.
(2010) presented HST/WFPC2 observations across the disk of NGC404 and
studied its star formation history in detail.

{\em AGC112521.}
This dIrr galaxy of low surface brightness  was detected in the
``blind'' HI survey ALFALFA, performing with the Arecibo radio
telescope (Saintonge et al. 2008). Our $H\alpha$ image of the
galaxy shows only one compact emission knot, which was
prevoiusly spotted by Zitrin \& Brosch (2008).

{\em KK13, KK14, KK15.} These are dwarf irregular companions of
the spiral galaxy NGC672, discovered by Karachentseva \&
Karachentsev (1998), and detected in the HI surveys by Huchtmeier
et al. (2000) and in the ALFALFA. All three objects have several
compact emission knots, which were also  noted  by Zitrin \& Brosch (2008).

{\em IC1727, NGC672.}
These are a pair of spiral galaxies of late types (Sm, Sd). Their distance,
 7.2 Mpc, was estimated by I.~Drozdovsky (private
communication) from the luminosity of brightest stars. It is the
distance we ascribed to the four remaining members of the NGC672
group: AGC112521, KK13, KK14 and KK15. $H\alpha$ images of both
spirals exhibit many HII regions, typical for Sm and Sd galaxies.
Integrated $H\alpha$ fluxes of IC1727 and NGC672 are in good
agreement with the fluxes, previously measured by Kennicutt et al.
(2008), but are nevertheless significantly (by two to three times) higher
than the fluxes, given by Zitrin \& Brosch (2008).

{\em UGC1281, KK16, KK17.}
These three dwarf galaxies form a group along with a brighter Sd
galaxy NGC784. Their distances  were measured by Tully et al.
(2006) using the TRGB method. Zitrin \& Brosch (2008) imaged all
 four galaxies in $H\alpha$. Their descriptions of emission
components in these galaxies are consistent with what we see in
Figure 3. However, in the case of KK16, where only a faint diffuse
emission is visible, the $H\alpha$ flux  measured by us is five
times higher than that obtained by Zitrin \& Brosch (2008).
 A reason for this difference remains unclear to us.

{\em UGC1703 = KKH9.}
This is a dwarf spheroidal galaxy at a distance of $4.2\pm0.3$ Mpc
estimated by Rekola et al. (2005) from the fluctuations of its
surface brightness. Judging from this distance, UGC1703 may either
be associated with the far periphery of the dwarf galaxy group
around NGC784, or belong to the rare class of isolated dSph
galaxies. Our $H\alpha$ image of UGC1703 does not show any signs
of emission. Table 3 indicates the upper limit of its $H\alpha$
flux.

{\em NGC855.}
This isolated elliptical galaxy is similar to NGC404 in terms of
luminosity and hydrogen mass. Its distance, 9.73 Mpc, was
determined by Tonry et al. (2001) from  fluctuations of surface
brightness. Wallington et al. (1988) noted the presence of a
ring-shaped HI shell around it, which is atypical for E galaxies. Our
snapshot of NGC855 in the $H\alpha$ line reveals bright emission in the
circumnuclear area, external parts of which are extended in the
polar directions. To understand the nature of this object with
hybrid properties of E and S galaxies, it is necessary to
investigate its kinematics in the HI and $H\alpha$ lines.

{\em AGC122226= KUG0243+275.}
This isolated blue compact galaxy was detected in the HI line in
the Arecibo survey (Saintonge et al. 2008). The $H\alpha$ image
shows active star formation, concentrated in several knots within the
central region of this galaxy.

{\em AGES.}
This dIrr galaxy was discovered as a result of the ``blind'' AGES
survey in the HI line of the vicinity of an isolated galaxy
NGC1156 (Minchin et al. 2011). Judging from its radial velocity,
the AGES object is a dwarf companion of NGC1156 at the projection
distance of $\sim80$ kpc, if we adopt for it the same
distance as that of NGC1156, 7.8 Mpc. Our image of AGES in the
$H\alpha$ line reveals two diffuse HII regions with a total star
formation rate of $\sim1\times
10^{-3}M_{\odot}$/year.

{\em KKH18.}
KKH18 is an isolated dIrr galaxy, the distance to which, 4.43 Mpc, is
determined via the TRGB (Karachentsev et al. 2003b). KKH18 and
UGC1703 are possibly forming the eastern extension of a filament
of dwarf galaxies, the dense part of which is  also hosting the
group around NGC784. A compact HII region and weak diffuse
emission in the center are visible on the body of KKH18.

{\em UGC2773.}
This is an isolated BCD galaxy in the region  of significant
($A_B=2.43^m$) absorption. Its $H\alpha$ image demonstrates many
compact HII regions, as well as a considerable diffuse emission.
 The distance to UGC2773 is estimated by us simply from the
radial velocity taking the Hubble parameter of $H_0=72$ km
s$^{-1}$Mpc$^{-1}$. Of course, in the local universe, peculiar
motions can dominate over the systematic Hubble component. However,
at present there is no generally accepted model that describes
the local peculiar velocity field well right taking into account the Virgocentric
infall and the Local Velocity Anomaly (see details in Tully et al.
2008).

{\em UGC2905.}
UGC2905 is an isolated dIrr galaxy on the southern part of which a background
spiral neighbor is projected. The distance to UGC2905, 5.8 Mpc, is
estimated from the brightest stars (Georgiev et al. 1997). Its
$H\alpha$ image reveals several compact and diffuse HII regions.

{\em UGC 3303.}
This is an isolated Sm galaxy with a bright star projected on its
central part. The distance of UGC~3303 is estimated as 7.2 Mpc
from the brightest stars (Makarova \& Karachentsev 1998). It may
be located in the periphery of a scattered association of
galaxies, the brightest member of which is the Orion galaxy. The
$H\alpha$ image reveals a lot of small HII regions scattered
across the disk of the galaxy.

{\em KK49 = CGCG422--003.}
This is a BCD galaxy in the Orion complex. Its distance is
evaluated from the radial velocity. The $H\alpha$ image of the KK49
body looks granulated because of the tightly located emission
knots.

{\em Orion, An0554.}
These are two galaxies (Sm and dIrr) located in the Orion complex in the
region of significant galactic absorption. Their distances
(6.4 and 5.5 Mpc, respectively) are determined by Karachentsev \&
Musella (1996) from the luminosity of brightest stars. The HII
regions, more abundant in the Orion galaxy in accordance with its
morphological Sm type, are visible on their $H\alpha$ images.
Recently, Cannon et al. (2010) carried out $H\alpha$ and VLA HI
observations of the Orion galaxy and found the rotating HI disk
extending far outside the optical boundary of the galaxy.

{\em HIZOA J0630+08.}
This HI source detected in the survey by Donley et al. (2005) is
located in a dense stellar region of the Milky Way at the galactic
latitude $b=-0.9^{\circ}$. On the POSS-II blue and red images, not a
single galaxy is seen within the radio telescope beam
($\sim15^{\prime}$). Our $H\alpha$ image does not show an
optical counterpart to this radio source either, which is most
likely a dIrr galaxy of low surface brightness, weakened by
absorption ($A_B=2.95^m)$.

{\em UGC3476, UGC3600, UGC3698.}
These are three  isolated dIrr galaxies, the distances to which are found
from the brightest stars (Makarova \& Karachentsev 1998). Each one
of them demonstrates the presence of active star formation sites
which is characteristic of isolated irregular galaxies.

{\em UGC3755.}
This is a dIrr galaxy, the distance to which is measured by Tully
et al. (2006) applying the TRGB. Its image in $H\alpha$ indicates
active star formation, most pronounced in the western part of the
galaxy.

{\em  DDO47, KK65 = CGCG087--033.}
These two galaxies are dwarf galaxies, forming an isolated pair
with a difference
in radial velocities of only 6 km s$^{-1}$.  The distances
measured by Tully et al. (2006) via the TRGB confirmed a physical
relationship of these galaxies. In both galaxies there are visible
compact emission regions.

{\em KK69, KK70.}
KK69 and KK70 are two dwarf companions of the spiral galaxy NGC2683. Both have low
surface brightness. An irregular dwarf, KK69, is characterized by a
very narrow HI emission  with a line width of 16.5$\pm$0.6 km
s$^{-1}$ (Huchtmeier et al. 2003). Our $H\alpha$ image shows a
very red or emission star-like object within its optical
boundaries. The nature of this object can be clarified by spectral
observations. The dwarf spheroidal system  KK70 lacks any signs of
$H\alpha$ emission.

{\em NGC2787, NGC4600.}
These are two isolated lenticular galaxies, the distances to which are
determined from surface brightness fluctuations (Tonry et al.
2001). In both cases the central parts of galaxies are
over-exposed, which makes the assessment of the flux in $H\alpha$
somewhat uncertain.

\section{External comparison of $H\alpha$ fluxes}

The accuracy of measurement of the $H\alpha$ flux of a galaxy, and of
the SFR value determined from it, depends on many factors. If
 variable atmospheric conditions were successfully monitored
during the observations by regular calibration  using the
spectrophotometric standards, the main source of errors for the
$F(H\alpha)$ values is inaccurate subtraction of the sky
background on the obtained images. For compact starburst galaxies
these errors are negligible, but for the galaxies with weak and
diffuse $H\alpha$ emission, such errors appear to reach $\sim10 --
20$\%. Note also that some authors determine the integrated flux of
a galaxy as the sum of its separate HII regions, while others also
take into account the general diffuse component, which also
gives rise to differences in the data they provide.

We evaluated our typical accuracy of $F(H\alpha)$ as
$\sim15$\%, or $\pm0.06$ dex in the logarithmic scale. However,
this internal assessment needs to be subject to an independent
external audit.

Among the 52 galaxies we observed there are 12 objects, in which
the $H\alpha$ fluxes were measured by Kennicutt et al. (2008), and
11 galaxies observed in $H\alpha$ by Zitrin \& Brosch (2008). The
data on $H\alpha$ fluxes in these galaxies are presented in Table
4. In the case of Kennicutt et al. (2008), we also cite the
internal flux measurement errors indicated by them.

A comparison of our $\lg F(H\alpha)$ values with the data by
Kennicutt et al. (2008) yields the mean square difference
$\sigma(\Delta\lg F)$= 0.09 and the average difference $\langle\lg
F_{KK} - \lg F_{Ken}\rangle=+0.004\pm0.03$, which indicates a good
agreement of independent measurements. The internal error of our
measurements that we have estimated as $\sigma(\lg F)=0.06$ is
approximately the same as in Kennicutt et al. (2008) (0.058), and
their quadratic sum reproduces well the mean square difference
$\sigma(\Delta\lg F)$ = 0.09. Note, however, that the agreement of
our data with the $H\alpha$ fluxes, published by Zitrin \& Brosch
(2008) turned out to be much worse: $\sigma(\Delta\lg F)=0.34$ ¨
and $\langle\lg F_{KK} - \lg F_{NB}\rangle=+0.10\pm0.11$.

A transition from the measured $H\alpha$ flux of a galaxy to the
SFR value is accompanied by additional errors, which are usually
systematic. These factors include: the contribution of the
emission line [NII] in the total registered $H\alpha$ + [NII]
flux, different methods of correction for internal absorption in
a galaxy, uncertainty of the Galactic absorption value according
to Schlegel et al. (1998) at low latitudes, underestimation of the
diffuse emission component of very low surface brightness, and
underestimation of possible HII regions in the distant periphery
of a galaxy (the case of NGC404). Finally, as noted above, the
transformation of $F(H\alpha)$ into the SFR via the linear
relationship (3) can significantly (by one to two orders of magnitude)
underestimate the true star formation rate due to the simplistic
notions on the initial stellar mass function in dwarf systems
(Pflamm-Altenburg \& Kroupa 2009).

\section{Some basic scaling relations}

Estimates of the global star formation rate are currently
obtained for 435 LV galaxies. As it is noted by many authors
(Karachentsev \& Kaisin 2007; James et al. 2008; Thilker et al.
2007; Lee et al. 2009) that SFR value correlates with the
integrated luminosity of a galaxy, its morphological
type, color index and hydrogen mass. The data on the dependence of
an SFR of a galaxy on its environment are rather contradictory
(Hunter \& Elmegreen 2004; James et al. 2004), but the prevailing
view is that such a dependence, if it exists, is weak, i.e. the
process of star formation in the galaxy is more likely driven
by its internal state than by external factors. Nevertheless,
there are well-known cases where a close interaction or merger of
galaxies leads to a spectacular burst of star formation or, the
other way around, a passage of a dIrr galaxy close to a massive
spiral suppresses star formation in the dwarf system due to
gas stripping from its shallow potential well.

Figure 4 represents a relation between the global star formation
rate and blue absolute magnitude for 435 LV galaxies. The
galaxies of different morphological types are shown by characters
of different colors. Empty symbols with arrows mark the instances
when only the upper limit of the SFR of a galaxy is known, determined
from Equation (3). The straight line in the figure corresponds to
the constant specific SFR per luminosity unit, {\bf $ SSFR =
0.70 \times10^{-10}M_{\odot} yr^{-1}L_{\odot}^{-1}.$}
Evidently, dwarf galaxies are
systematically located below this line. According to
Pflamm-Altenburg \& Kroupa (2009) their deviation from the linear
relation is leveled if the transition from the measured
$H\alpha$ flux to the SFR is made in the light of modern ideas on
the initial stellar mass function in dwarf systems. The
galaxies from Tables 2 and 3 do not noticeably stand out among the
rest of the objects. A distinctive feature of the \{$SFR, M_B$\}
diagram is the presence of a rather sharp upper boundary,
$ SSFR = 4.3 \times10^{-10}M_{\odot} yr^{-1}L_{\odot}^{-1}$,
which is mainly traced by dIrr, BCD, and Sm--Sc galaxies.
Of the galaxies listed in Tables 2 and 3, UGC2773, KK35 and
NGC1569 present examples of such cases. The existence of
a critical upper value for the  SSFR is obviously an
important universal parameter, characterizing the process of
conversion of gas into stars.

The distribution of LV galaxies by SFR values and hydrogen mass is
presented in Figure 5, where the symbol designations are the same
as in Figure 4. The dashed line corresponds to the constant specific
SFR, related to the unit of hydrogen mass. The solid line shows a
steeper dependence of SFR $\propto M^{3/2}_{HI}$, followed by
individual emission complexes inside the galaxies (the
Schmidt--Kennicutt law). According to Pflamm--Altenburg \& Kroupa
(2009), recalculation of SFR values for dwarf galaxies with the
view of a more accurate initial stellar mass function decreases
the regression slope in Figure 5 from 1.5 to 1.0.

To describe the evolutionary status of various samples of
galaxies, Karachentsev \& Kaisin (2007) suggested to use the
diagnostic "past--future" diagram, where the dimensionless and
distance-independent parameters

$$P=\log\{[{\rm SFR}]\times T_0/L_K\}, \,\,\,F=\log\{1.85M_{HI}/([{\rm SFR}]\times T_0)\}\eqno(4)$$

show which part of the observed stellar mass of the galaxy can be
reproduced at the now observed SFR during the cosmological time
$T_0$, and for how long the star formation can continue there with
the present gas reserves of $M_{gas}=1.85 M_{HI}$. Here the factor
1.85 gives a correction for the average abundance of helium and molecular
gas in the galaxy (Fukugita \& Peebles, 2004). For the P parameter in
expression (4), we use a known fact that the infrared K-band
luminosity $L_K$ of a galaxy reproduces its stellar mass at
$M_*/L_K=1M_{\odot}/L_{\odot}$ (Bell et al. 2003; Karachentsev \&
Kut'kin 2005). We adopted  $K_s$-band magnitudes for 122 LV galaxies
from the 2MASS survey (Jarrett et al. 2003). For the remaining objects
we transferred their B-magnitudes into the $K_s$ ones, using the empirical
relations between the average color index $\langle B-K\rangle$ and the
morphological type of a galaxy, discussed by Jarrett et al. (2003) and
Karachentsev \& Kut'kin (2005):

\[\begin{array}{llll}
\langle B-K\rangle=4.10     &{\rm for} & T\leq2 & \\
\langle B-K\rangle=4.60-T/4 & {\rm for} &T=3-8 & \,\,\,\,\,\,\,\,\,\,\,\,\,\,    (5)\\
\langle B-K\rangle=2.35     & {\rm for} &T=9,10.  & \\
\end{array}\]

The distribution of 435 LV galaxies on the "Past--Future" plane is
presented in Figure 6, where galaxies of different types, (E, S0, dSph),
(Sa, Sab, Sb, Sbc), (Sc, Scd, Sd, Sdm, Sm), and (Irr, BCD), are given in four
separate panels. As above, open symbols with arrows indicate objects
with only the upper limit of SFR or HI-flux.  Here, we omitted 41
galaxies with the upper limit of both SFR and HI-flux because of their
uncertain position on the F--scale.

It is easy to see that the galaxies of different morphological types
occupy different regions on the $\{P,F\}$ plane, demonstrating the
expected evolutionary segregation. The evolutionary trend
according to galaxy types is also reflected in the data of Table 5. Its
columns indicate: (1) --- morphological type, (2) --- number of
galaxies of this type in the LV with measured SFRs, (3,4) --- median
values of the $P$ and $F$ parameters. We can draw the following
conclusions from these data.
\begin{enumerate}

\item The current SFRs in the E, S0, and dSph galaxies
can reproduce only about 2\% of their stellar mass, therefore in
the past their average SFR was significantly
higher. Typical gas reserves in the E, S0, and dSph galaxies are
rather uncertain, and their typical gas consumption timescale remains
uncertain too.

\item According to the median parameters $P$ and $F$, the spiral
galaxies of early types, dominated by the bulges, have already
passed the peak of their evolution. The past SFR
was an order of magnitude higher than the present one, and the
current gas reserves can support the SFRs during merely 28\%
of the cosmological timescale.

\item In disk-like galaxies of the late Sc--Sm types, the current
SFR is only slightly lower than it was in the
past. The resources of gas in Sc--Sdm galaxies are supplying their
observed SFRs during almost another Hubble time.

\item The population of Irr and BCD galaxies had almost the same mean
SFR in the past, as it does now. Their gas reserves are
sufficient for further star formation on a timescale of around
$1.8 T_0$. The diagonal character of the distribution of these
galaxies on the  $\{P, F\}$ plane obviously points to the
variability of SFR in galaxies of low masses.
Facing periodic bursts, dIrr galaxies are moving from the top left
to the bottom right quadrant, acquiring the signs of BCD galaxies.
Note that Stinson et al. (2007) simulated the evolution of dIrr
galaxies taking into account effects of gas outflow due to the
wind from SNe, and found cyclic bursts of star formation on the
scale of $\sim0.3$ Gyr with an amplitude of
$\sim(2-3)^m$ for dwarf systems of very low masses.

\item Scattering of the LV galaxies on the $\{P, F\}$ diagram  is
quite high, reaching two to four orders of magnitude depending on the
morphological type. As we already noted, the $H\alpha$ flux
measurement error normally does not exceed $\sim0.1$ dex,
although near the detection limit these errors can be much higher.
The uncertainty of transformation of $F(H\alpha)$ into [SFR],
discussed by Pflamm--Altenburg \& Kroupa (2009) also affects
the parameter spread, but it shifts the galaxies exactly arriswise
$F=-P$. Thus, much of the galaxy dispersion in Figure 6 does not have
an instrumental origin, but a cosmic one. The smallest dispersion,
$\sigma(P)=0.4,\sigma(F)=0.6$, is observed for the population
of late type-spirals, Sc--Sdm. It is most likely that the rotation
of Sc--Sdm galaxies and the stimulation of star formation  it causes in gas-rich disks
makes this process fairly regular.
\end{enumerate}

\section{The present cosmic SFR density}

As demonstrated by Madau et al. (1996), Villar et al. (2008), Gonzalez
et al. (2010), Westra et al. (2010), and other authors, the average
SFR in previous epochs $z\simeq 1-2$ was an order
of magnitude higher than nowadays. Analyzing the change in the
average SFR density from redshift
$\rho_{SFR}(z)$,  it is important to reliably fix the current
value of $\rho_{SFR}(0)$ from the observations of nearby galaxies.

To this end, we used all available data on the SFR of galaxies
situated within 8 Mpc at galactic latitudes $|b| > 10^{\circ}$.
We did not
consider more distant objects because the present completeness of
the $H\alpha$ survey drops appreciably beyond 8 Mpc. The integrated
SFR for the 8 Mpc sample amounts to 53 $M_{\odot}yr^{-1}$. As  is
seen from Figure 1, the present $H\alpha$ survey is quite complete up
to $M_B = - 15^m$. Based on the relation "SFR versus $M_B$" in Figure 4,
we estimate that the integrated contribution of dwarf galaxies still
unobserved in $H\alpha$ adds about 4 $M_{\odot}yr^{-1}$ to the total amount.
Therefore, the mean SFR density within 8 Mpc turns
out to be $\rho_{SFR}(0)= 0.032 M_{\odot} yr^{-1} Mpc^{-3}$. As  was
noted by Karachentsev \& Kut'kin (2005), the mean stellar mass
density within 8 Mpc, estimated from the K-band luminosity density
$ j_K(L\mid 8 Mpc) = 6.8\times10^8L_{\odot} Mpc^{-3}$, appeared to be
$(1.7\pm0.2)$ times higher than the mean cosmic density
$j_K(L)_{cosmic}=(3.8\pm0.6)\times10^8L_{\odot} Mpc^{-3}$ obtained by
Cole et al. (2001) and Bell et al. (2003) from  2MASS.
Reducing for the local overdensity,  yields  the mean cosmic
density of SFR in the present epoch

$\rho_{SFR}(0)=(0.019\pm0.003)M_{\odot} yr^{-1}Mpc^{-3}$.

Table 6 gives a comparison of our estimate with the data obtained
by other authors based on the samples of different depths and
different compilation methods. As one can see, the agreement of
independent estimates of $ \rho_{SFR}(0)$ is quite satisfactory.

The data from Table 7, gathering the values of some basic cosmic
parameters describing the star formation within 1 Mpc$^3$ at z=0
and h=0.72 can be useful to validate various models of galaxy
evolution.  The rows of the table present: (1) --- the critical density
of matter, (2) --- the luminosity density in the $K$-band (also
evaluating the mean density of stellar mass at $M_*/L_K = 1
M_{\odot}/L_{\odot}$), (3) --- the mean density of hydrogen mass
according to HIPASS (Zwaan et al. 2003), (4) --- the mean density of
SFR, and (5,6) --- the mean density of the dimensionless
parameters $P$ and $F$, derived from the quantities of
rows (2)--(4) via Equations (4). The value  $\rho(P)=-0.17$, actually
averaged with the galaxy masses proportional to their
$K$--luminosity, means that the current star formation rate in a
unit volume is only 1.5 times lower than the average SFR
in past epochs. This result looks significantly at odds with
the notion that the characteristic SFR in the
$z\simeq1$ era was an order of magnitude more intense than at
$z=0$. The other value, $\rho(F)=-0.50$, means that an average unit
volume has such a reserve of gas in the galaxies, which is able to
maintain the average present rate of star formation in them for
another 4--5 Gyr. In other words, our universe has already gone more
than halfway in the history of transformation of gas into stars
and is now being on the descending branch of this process
immediately after the era of peak intensity of star formation. It
is needless to stress that this assertion is true only under the
condition that the bulk of gas is located in the volume of the
galaxies themselves, rather than being distributed in the
intergalactic space.

\section{Concluding remarks}

The program of our massive $H\alpha$ survey of galaxies in the
neighboring volume a radius of 10 Mpc allows us to determine some basic
cosmic parameters, characterizing the rate and resource of star
formation in the local universe. An important prerequisite for
this is an exclusion of deliberate selection in the choice of
objects for the observational program by morphological type and/or
other features. A simple principle is evident here: the lower the
selectivity of objects is for observations, the simpler the
interpretation of the data obtained will be. The lack of accurate distance
measurements to a part of nearby galaxies somewhat blurs this
ideal situation. It should be noted, however, that positions of
galaxies in the diagnostic diagram \{P, F\} (Figure 6) do not depend
on the errors of distance finding.

Keeping  the E, S0, and dSph galaxies, which are not
expected to have  $H\alpha$ emission, in our sample we found surprisingly
that in some of them the process of star formation goes on at a
fairly high rate. For example, spheroidal galaxies KDG61, DDO44
and KKR25 indicate the presence of separate emission knots, in
which a young stellar population is formed. The isolated E and S0
galaxies, NGC404, NGC855, and NGC4460, demonstrate active $H\alpha$
emission in their central regions, which probably indicates a
constant inflow of the accreting intergalactic gas  into these
galaxies (Moiseev et al. 2010). There are reasons to assume
that a population of young semi-formed dwarf galaxies, similar to the
HI-clouds in the Virgo (122746.2 +013601) and CVnI (122043.4
+461233) clusters, or to the HIJASS  (102100.2 +684200) and Leib
(AGC219303) objects in the M81 and LeoI groups is located in the
upper right quadrant of the diagnostic diagram (Figure 6). To
clarify the nature of such objects with masses comparable to the
masses of dwarf galaxies, much deeper observations  are needed
with  flux limit of $F(H\alpha)\sim 10^{-16}$ erg/cm$^2$sec and
$F(HI) \sim 10^{-2} Jy$ km/sec. Such observations would obviously
require a significant amount of time on large telescopes. This demand
should be related to the predictions of modern scenarios of galaxy
evolution. So far, the existing shallow $H\alpha$ survey of
nearby dwarf galaxies is successfully competing with another
survey of these galaxies in the ultraviolet range on GALEX
(Gil de Paz et al. 2003) due to weaker absorption in the $H\alpha$
line and higher angular resolution.

\acknowledgements{The work was supported by the Russian
Foundation for Basic Research (projects 10--02--00123,
09--02--90414-UKR-f-a and 10--02--92650-IND-a).

{\bf References}  \\
\par\noindent Abazajian K., et al. 2009, ApJS, 182, 543
\par\noindent Afanasiev V.L., Gazhur E.B., Zhelenkov S.R. \& Moiseev A.V. 2005,
 Bull.SAO, 58, 90
\par\noindent Bell E.F., McIntosh D.H., Katz N., Weinberg M.D., 2003, ApJS, 149, 289
\par\noindent Bell E.F. \& Kennicutt R.C. 2001, ApJ, 548, 681
\par\noindent Bouchard A., Da Costa G.S., Jerjen H.,  2009  AJ, 137, 3038
\par\noindent Brinchmann J., Charlot S., White S.D.M., et al, 2004, MNRAS, 351, 1151
\par\noindent Cannon L.M., Haynes K., Most H. et al. 2010, AJ, 139, 2170
\par\noindent Cole S., et al. 2001, MNRAS, 326, 255
\par\noindent Colless M., et al. 2001, MNRAS, 328, 1039
\par\noindent Del Rio M.S., Brinks  E., Cepa  J.  2004, AJ, 128, 89
\par\noindent Donley J.L., Staveley-Smith L., Kraan-Korteweg R.C., et al, 2005, AJ, 129, 220
\par\noindent Epinat B., Amram P., Marcelin M., 2008, MNRAS, 390, 466
\par\noindent Fingerhut R.L., McCall M.L., De Robertis M., et al, 2003, ApJ, 587, 672
\par\noindent Fukugita M., Peebles P.J.E., 2004, ApJ, 616, 643
\par\noindent Gallagher J.S., Hunter D.A. \& Tutukov A.V. 1984, ApJ, 284, 544
\par\noindent Gallego J., Zamorano J., Aragon-Salamanca A., Rego M., 1995, ApJ, 455, 1
\par\noindent Georgiev Ts.B., Karachentsev I.D., Tikhonov N.A., 1997, Astron. Lett., 23, 514
\par\noindent Gil de Paz A., Madore B.F. \& Pevunova O. 2003, ApJS, 147, 29
\par\noindent Giovanelli R., et al. 2005, AJ, 130, 2598
\par\noindent Gonzalez V., Labbe I., Bouwens R.J. et al, 2010, ApJ, 713, 115
\par\noindent Grocholski A.J., Aloisi A., van der Marel R.P., et al, 2008, ApJ, 686L, 79
\par\noindent Hanish D.J., Meurer G.R., Ferguson H.C., et al, 2006, ApJ, 649, 150
\par\noindent Huchtmeier W.K., Karachentsev I.D. \&  Karachentseva V.E., 2003, A\&A, 401, 483
\par\noindent Huchtmeier W.K., Karachentsev I.D. \&  Karachentseva V.E., 2001, A\&A, 377, 801
\par\noindent Huchtmeier W.K., Karachentsev I.D., Karachentseva V.E. \& Ehle M.
     2000, A\&AS, 141, 469
\par\noindent Hunter D.A. \& Elmegreen B.G. 2004, AJ, 128, 2170
\par\noindent Hunter et al. 1993  AJ, 106, 1797
\par\noindent James P.A., Knapen J.H., Shane N.S., et al., 2008, A\&A, 482, 507
\par\noindent James P.A., Shane N.S., Beckman J.E., et al. 2004, A\&A, 414, 23
\par\noindent Jarrett T., Chester R., Cutri R., et al, 2003, AJ, 125, 525
\par\noindent Jones D.H., et al. 2004, MNRAS, 355, 747
\par\noindent Kaisin  S.S.,\& Karachentsev  I.D., 2008, A\&A, 479, 603
\par\noindent Kaisin  S.S., Karachentsev  I.D., 2006, Astrofizika, 49, 287
\par\noindent Kaisin  S.S., Kasparova A.V., Kniazev A.Yu., Karachentsev I.D., 2007, Astron. Lett., 33, 1
\par\noindent Karachentsev I.D., 1994, Astron. Astrophys. Trans. 6, 1
\par\noindent Karachentsev I.D., Karachentseva V.E., Huchtmeier W.K., Makarov D.I., 2004,
 AJ, 127, 2031 (= CNG)
\par\noindent Karachentsev I.D., Kaisin S.S., Tsvetanov Z., Ford H., 2005, A\&A, 434, 935
\par\noindent Karachentsev  I.D., Kaisin  S.S., 2007, AJ, 133, 1883
\par\noindent Karachentsev I.D., Sharina M.E., Dolphin A.E., Grebel E.K., 2003a, A\&A, 408, 111
\par\noindent Karachentsev I.D., et al, 2003b, A\&A, 398, 479
\par\noindent Karachentsev I.D., Dolphin A.E., Tully R.B., et al, 2006, AJ, 131, 1361
\par\noindent Karachentsev I.D. \& Kut'kin A.M., 2005, Astronomy Lett., 31, 299
\par\noindent Karachentsev I.D. \& Musella I., 1996, A\&A, 315, 348
\par\noindent Karachentseva V.E. \& Karachentsev I.D., 2000, A\&AS, 146, 359
\par\noindent Karachentseva V.E. \& Karachentsev I.D., 1999, A\&AS, 135, 221
\par\noindent Karachentseva V.E.,  Karachentsev I.D., \& Richter G.M., 1998, A\&AS, 127, 409
\par\noindent Kennicutt R.C., Lee J.C., Funes J.G., et al. 2008, ApJS, 178, 247
\par\noindent Kennicutt R.C. \& Kent S.M., 1983  AJ, 88, 1094
\par\noindent Kennicutt R.C. et al.,  1989  ApJ, 337, 761
\par\noindent Kraan-Korteweg R.C., Tammann G.A., 1979, Astron. Nachr., 300, 181
\par\noindent Lee J.C., Kennicutt R.C., Funes J.G., et al. 2009, ApJ, 692, 1305
\par\noindent Madau P., Ferguson H.C., Dickinson M.E., et al. 1996, MNRAS, 283, 1388
\par\noindent Makarova L.N. \& Karachentsev I.D., 1998, A\&AS, 133, 181
\par\noindent Meurer G.R. et al.  2006,  ApJS, 165, 307
\par\noindent Miller B.W. \& Hodge P. 1994, ApJ, 427, 656
\par\noindent Minchin R. F., Momjian E., Auld R., et al, 2011, AJ, accepted
\par\noindent Moiseev A.V., Karachentsev I.D. \& Kaisin S.S., 2010, MNRAS, 403, 1849
\par\noindent Oke J.B. 1990, AJ, 99, 1621
\par\noindent Perez-Gonzalez P.G., Zamorano J., Gallego J., et al, 2003, ApJ, 591, 827
\par\noindent Pflamm-Altenburg J., Weidner C., Kroupa P., 2007, ApJ, 671, 1550
\par\noindent Pflamm-Altenburg Kroupa P., 2009, ApJ, 706, 516
\par\noindent Rekola R., Jerjen H., Flynn C., 2005, A\&A, 437, 823
\par\noindent Saintonge A., Giovanelli R., Haynes M.P., et al, 2008, AJ, 135, 588
\par\noindent Salim S., Rich R.M., Charlot S., et al., 2007, ApJS, 173, 267
\par\noindent Schlegel D.J., Finkbeiner D.P. \& Davis M. 1998, ApJ, 500, 525
\par\noindent Spergel D.N., et al. 2007, ApJS, 170, 377
\par\noindent Stinson G.S., Dalcanton J.J., Quinn T., et al, 2007, ApJ, 667, 170
\par\noindent Thilker D.A., Bianchi L., Meurer G., et al., 2007, ApJS, 173, 572
\par\noindent Thilker  D.A., Bianchi L., Schiminovich D. et al. 2010, ApJ, 714, L171
\par\noindent Tonry J.L., Dressler A.,  Blakeslee J.P., et al. 2001, ApJ, 546, 681
\par\noindent Tresse L., Maddox S.J., 1998, ApJ, 499, 112
\par\noindent Tully R.B., Shaya E.L., Karachentsev I.D. et al. 2008, ApJ, 676, 184
\par\noindent Tully R.B., Rizzi L., Dolphin A.E., et al. 2006, AJ, 132, 729
\par\noindent Tully R.B. \& Fisher J.R., 1977, A\&A, 54, 661
\par\noindent van Zee L. 2000, AJ, 119, 2757
\par\noindent Verheijen M.A.W., 2001, ApJ, 563, 694
\par\noindent Villar V., Gallego J., Perez-Gonzalez P.G., Pascual S., 2008, ApJ, 677, 169
\par\noindent Wallington S., Katz N., Gunn J.E., et al, 1988, BAAS, 20, 1038, 42.06
\par\noindent Westra E., Geller M.J., Kurtz M.J., et al, 2010, ApJ, 708, 534
\par\noindent Williams B.F., Dalcanton J.J., Gilbert K.M., et al. 2010, ApJ, 716, 71
\par\noindent Young J.S., Allen L., Kenney J.D. \& Rownd B. 1996, AJ, 112, 1903
\par\noindent Zitrin A., Brosch N., 2008, MNRAS, 390, 408
\par\noindent Zwaan M.A., Staveley-Smith L., Koribalski B.S., et al, 2003, AJ, 125, 2842

  \clearpage
\begin{figure}
 \epsscale{1.0}
\plotone{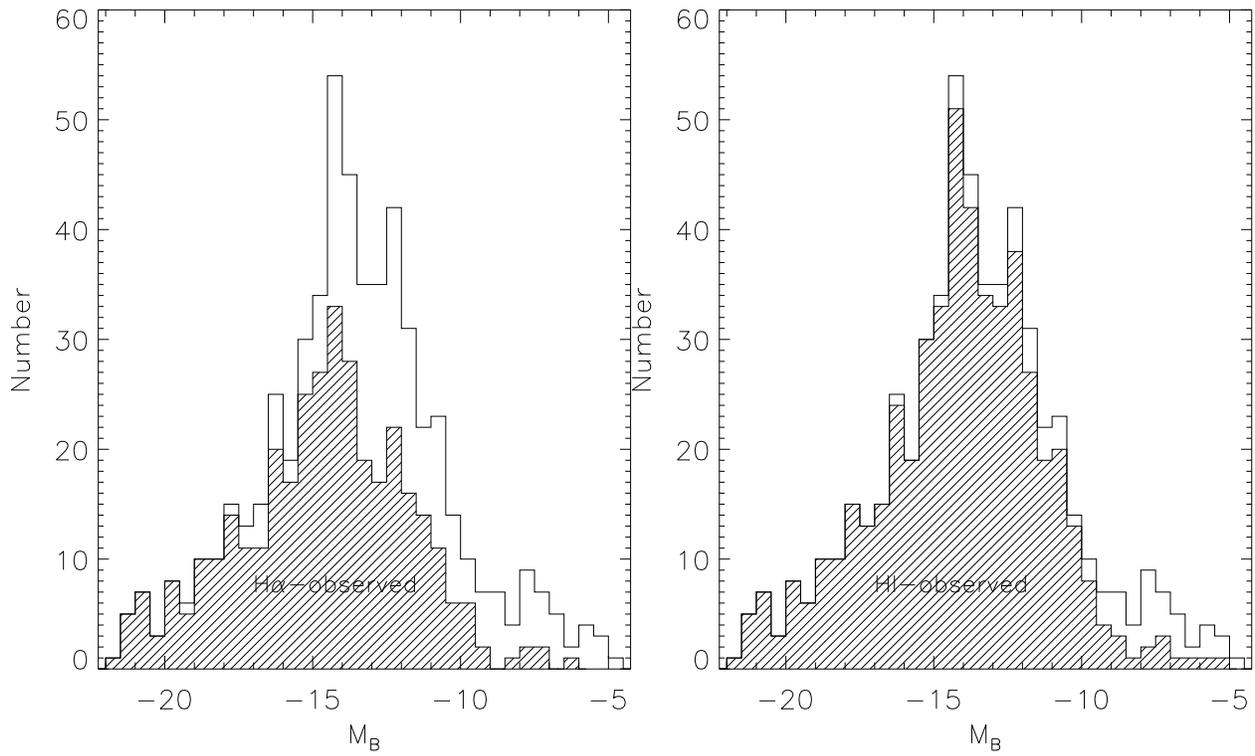}
\caption{ Histograms showing the fraction of the LV galaxies
observed in the H$\alpha$ line (left) and the HI-line (right) depending
on their blue absolute magnitude.}
\end{figure}

%

\begin{figure}
\figurenum{4}
\plotone{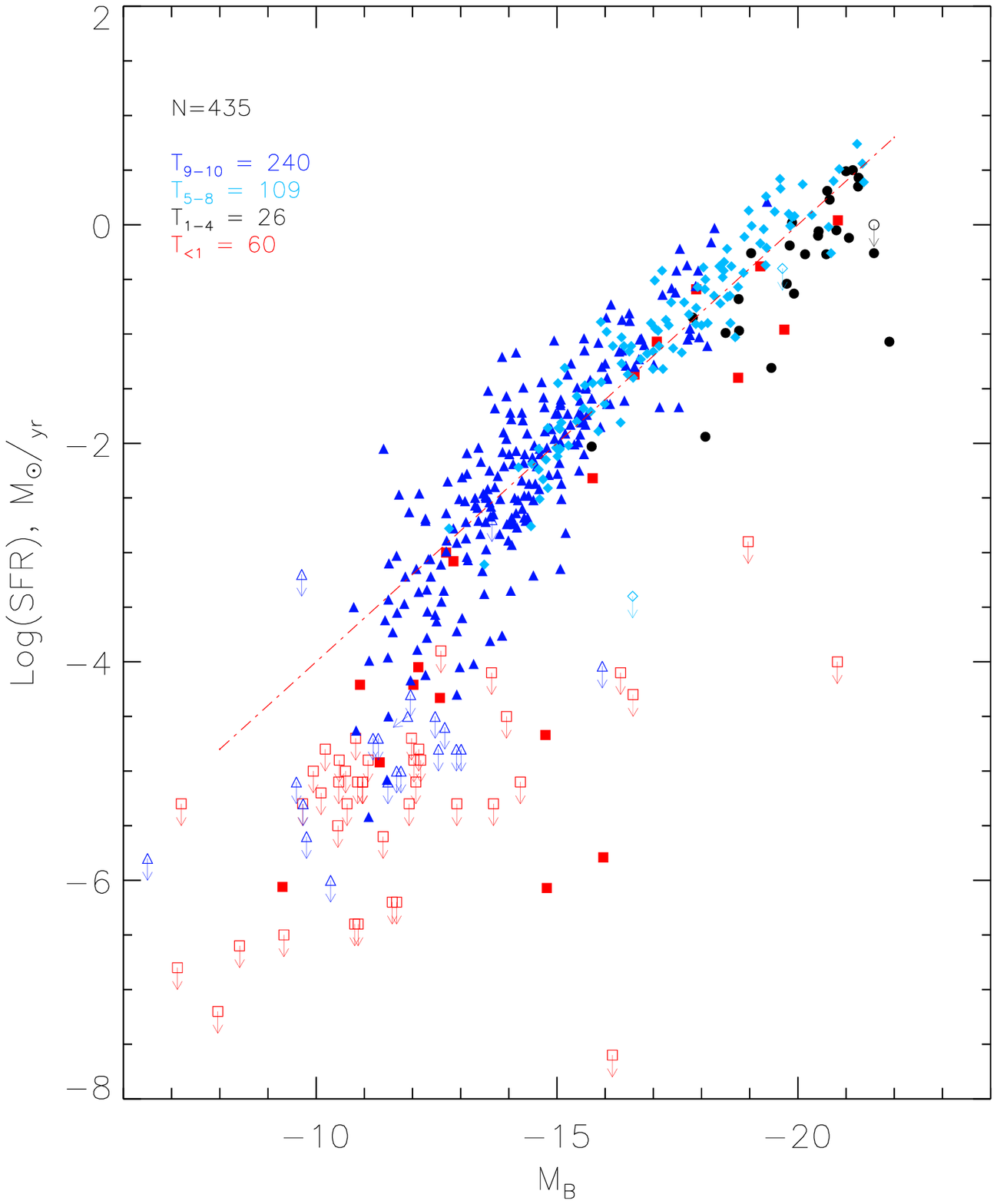}
\caption{SFR vs. blue absolute magnitude for
435 LV galaxies. The open symbols indicate the galaxies with
only the upper limit of their SFR. The line corresponds to a constant
SFR per unit luminosity.}
\end{figure}

\begin{figure}
\figurenum{5}
\plotone{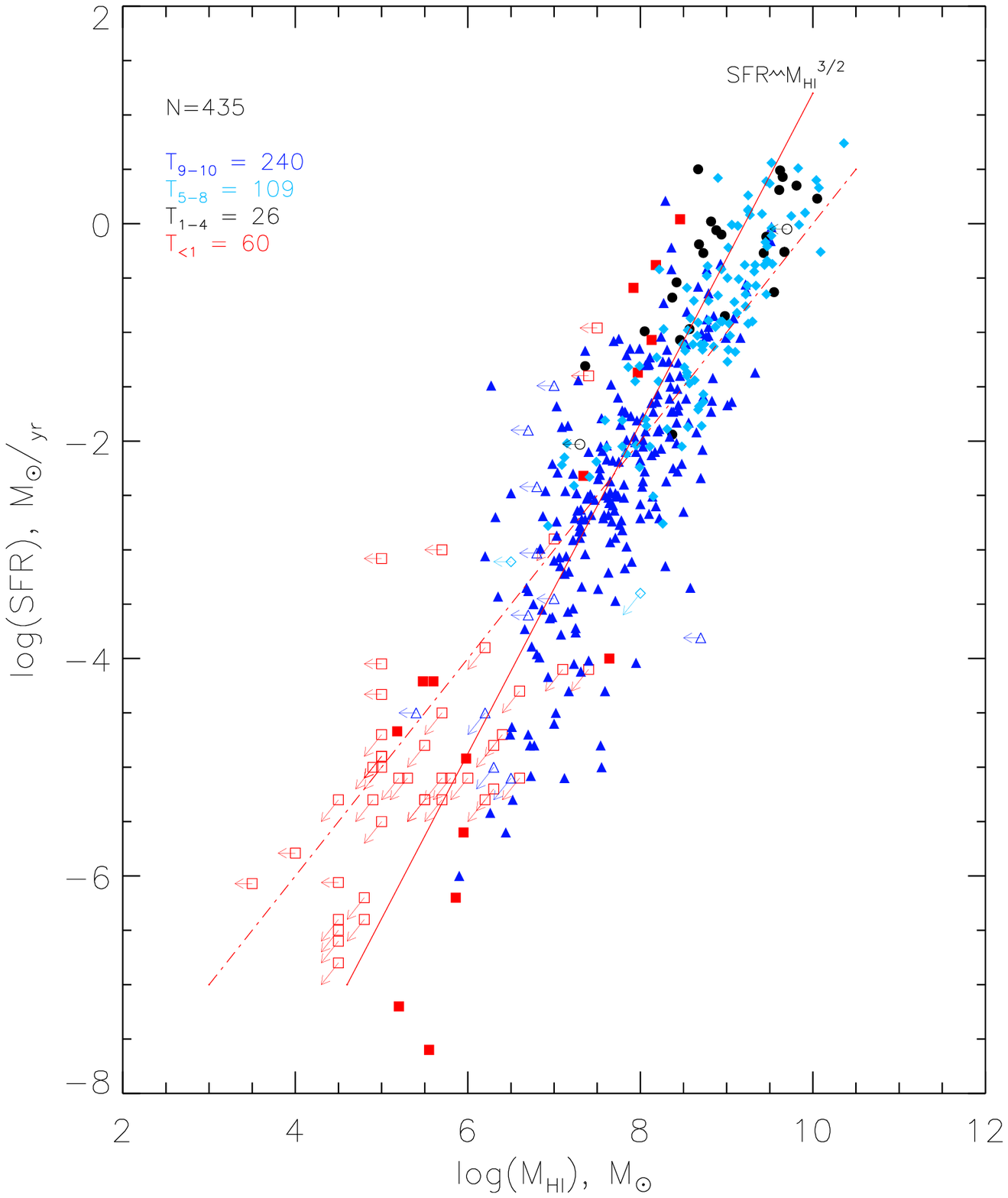}
\caption{SFR vs. neutral hydrogen mass for 435
LV galaxies. The objects with an upper limit of SFR or $M_{HI}$
are indicated by open symbols. The dashed line corresponds to a fixed $SFR$
per unit hydrogen mass and the solid line traces the relationship
SFR$\propto M_{HI}^{1.5}$.}
\end{figure}

\begin{figure}
\figurenum{6}
\plotone{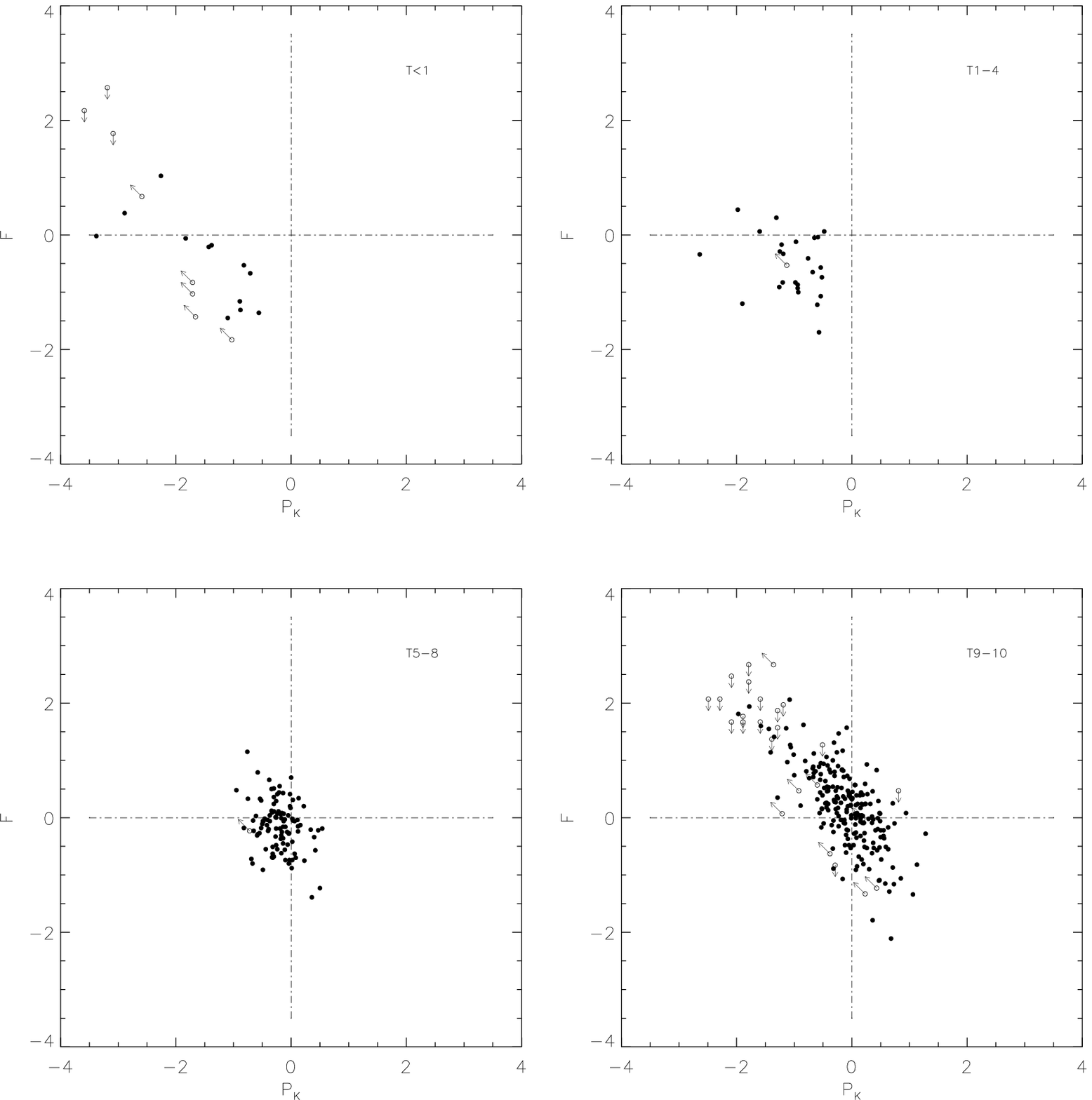}
\caption{LV galaxies of different morphological types
on the diagnostic diagram ``Past-Future''. The objects with an upper
limit of SFR or $M_{HI}$ are shown by open symbols with arrows.}
\end{figure}

\clearpage
\begin{table}
\caption{Basic contributions to the $H\alpha$ survey of the
LV galaxies}
\begin{tabular}{lcl} \hline
  Reference                  &  $N_{LV}$ &  Sample \\
\hline
  Kennicutt \& Kent, 1983     &   25  &  Spiral,Irregular   \\
  Kennicutt et al. 1989      &   14   & Spiral,Irregular   \\
  Hunter et al. 1993         &   37   & Irregular  \\
  Miller \& Hodge,1994        &   11  &  M81 group\\
  Young et al. 1996          &   16   & Spiral         \\
  van Zee, 2000              &   15   & Isolated irregular \\
  Bell \& Kennicutt, 2001     &   24  & Spiral,Irregular  \\
  Gil de Paz et al. 2003     &   10   & BCD        \\
  James et al. 2004          &   49   & S0/a--Im   \\
  Hunter \& Elmegreen, 2004   &   50  &  Im, BCD   \\
  Meurer et al. 2006         &   10   & HIPASS selected \\
  Epinat et al. 2008         &   27   & Spiral     \\
  Kennicutt et al. 2008      &  171   & T$>-1$, B $<15^m,\mid b\mid>20^{\circ}$ \\
  Bouchard et al. 2009       &   18   & Sculptor and CenA groups  \\
  This paper              &  207  &  All types   \\
\hline
\end{tabular}
\end{table}

\clearpage
\hoffset=-2cm
\begin{table}
\caption{Galaxies in the IC342/Maffei complex}
\begin{tabular}{lcrrlccccc} \hline
Name  &      RA    Dec   & $T$  &$V_{LG}$  &$D_{MW}$ &  $M_B$  &$\lg M(HI)$ & $\lg F(H\alpha)$ &$\lg Fc(H\alpha)$ & $\lg(SFR)$ \\
     &      (2000.0)    &    &km s$^{-1}$  & Mpc &  mag  & $M_{\odot}$  & erg/cm$^2$s       &erg/cm$^2$s   & $M_{\odot}$/yr \\
\hline
KKH5  &   010732.5+512625& 10 & 304  &4.26 &$-$12.27 & 6.87 &$-$13.31 &$-$13.05& $-$2.69 \\
KKH6  &   013451.6+520530& 10 & 270  &3.73 &$-$12.38 & 7.12 &$-$13.78 &$-$13.42& $-$3.18 \\
Cas1  &   020607.9+690036& 10 & 284  &3.3  &$-$16.70 & 8.11 &$-$12.32 &$-$11.37& $-$1.24 \\
KKH11 &   022435.0+560042& 10 & 308  &3.0  &$-$13.35 & 7.68 &$-$13.10 &$-$12.64& $-$2.59 \\
KKH12 &   022727.0+572916& 10 & 303  &3.0  &$-$13.03 & 7.53 &$-$13.11 &$-$12.37& $-$2.32 \\
MB1   &   023535.6+592247& 10 & 421  &3.0  &$-$14.81 & 7.23 &$-$13.37 &$-$12.46& $-$2.41 \\
Maffei1&  023635.5+593918& -3 & 246  &3.01 &$-$18.97 &  $-$   &$-$14.0: &$-$12.9:& $-$2.9: \\
Maffei2&  024154.5+593611&  4 & 212  &3.1  &$-$20.37 & 8.82 &$-$11.99 &$-$10.34& $-$0.35 \\
Dwing2&   025408.5+590019& 10 & 316  &3.0  &$-$14.55 & 8.30 &$-$13.52 &$-$12.42& $-$2.36 \\
MB3   &   025543.6+585142& 10 & 280  &3.0  &$-$13.65 & 6.32 &$-$14.0: &$-$12.8:& $-$2.7: \\
Dwing1&   025656.1+585442&  4 & 333  &3.0  &$-$18.93 & 8.63 &$-$12.22 &$-$10.81& $-$0.81 \\
KK35  &   034512.6+675150& 10 & 320  &3.16 &$-$14.30 & 6.27 &$-$12.13 &$-$11.59& $-$1.49 \\
UA86  &   035949.5+670731&  8 & 275  &2.96 &$-$17.95 & 8.79 &$-$11.94 &$-$11.06& $-$1.02 \\
CamA  &   042515.6+724821& 10 & 164  &3.93 &$-$14.06 & 8.22 &$-$13.19 &$-$12.99& $-$2.70 \\
N1569 &   043049.1+645053&  9 &  88  &3.36 &$-$19.36 & 8.29 &$-$10.59 &$-$9.94&  \,\,\,\,   0.21 \\
UA92  &   043200.3+633650& 10 &  89  &3.01 &$-$15.60 & 8.35 &$-$12.30 &$-$11.56& $-$1.51 \\
N1560 &   043249.9+715252&  7 & 171  &3.45 &$-$16.87 & 9.10 &$-$11.60 &$-$11.37& $-$1.19 \\
CamB  &   045306.9+670557& 10 & 266  &3.34 &$-$11.85 & 7.12 &$-$13.56 &$-$13.36& $-$3.21 \\
UA105 &   051415.1+623451&  8 & 279  &3.15 &$-$16.81 & 8.51 &$-$11.58 &$-$11.27& $-$1.17 \\
\hline
\end{tabular}
\end{table}

\begin{table}
\caption{Field galaxies observed in $H\alpha$ }
\begin{tabular}{lcrclccccc} \hline
Name  &    RA    Dec   & $T$  &$V_{LG}$ & $D_{MW}$ &  $M_B$  &$\lg M(HI)$ & $\lg F(H\alpha)$ &$\lg Fc(H\alpha)$ & $\lg SFR$ \\
     &     (2000.0)   &    &km s$^{-1}$ &  Mpc &  mag  & $M_{\odot}$  &erg/cm$^2$s     &erg/cm$^2$s    & $M_{\odot}$/yr \\
\hline
N404  & 010926.9+354303&  1 & 195 & 3.24 &$-$16.61 & 8.02 &$-$11.49 &$-$11.44& $-$1.32 \\
AGC112521 & 014107.9+271926& 10 & 482 & 7.2  &$-$11.26 & 6.89 &$-$14.81 &$-$14.74& $-$3.93 \\
KK13  & 014216.8+262204& 10 & 556 & 7.2  &$-$13.11 & 7.03 &$-$13.77 &$-$13.69& $-$2.87 \\
KK14  & 014442.7+271716& 10 & 622 & 7.2  &$-$12.13 & 7.51 &$-$14.24 &$-$14.17& $-$3.36 \\
KK15  & 014641.6+264805& 10 & 563 & 7.2  &$-$11.43 & 6.98 &$-$14.52 &$-$14.44& $-$3.63 \\
IC1727& 014730.1+271952&  8 & 535 & 7.2  &$-$17.71 & 9.16 &$-$11.94 &$-$11.84& $-$1.02 \\
N672  & 014753.2+272601&  7 & 618 & 7.2  &$-$18.76 & 9.24 &$-$11.56 &$-$11.39& $-$0.57 \\
U1281 & 014932.3+323533&  7 & 367 & 4.97 &$-$15.86 & 8.31 &$-$12.46 &$-$12.37& $-$1.88 \\
KK16  & 015520.6+275715& 10 & 400 & 5.40 &$-$12.65 & 6.68 &$-$13.97 &$-$13.91& $-$3.35 \\
KK17  & 020009.9+284957& 10 & 360 & 4.92 &$-$11.50 & 6.35 &$-$13.96 &$-$13.91& $-$3.43 \\
N784  & 020116.8+285037&  8 & 386 & 4.97 &$-$16.58 & 8.54 &$-$11.95 &$-$11.90& $-$1.40 \\
U1703 & 021255.8+324851& -2 &  -  & 4.2  &$-$11.54 & 6.3: &$-$15.2: &$-$15.1:& $-$4.7: \\
N855  & 021403.7+275238& -5 & 774 & 9.73 &$-$17.07 & 8.13 &$-$12.14 &$-$12.05& $-$0.97 \\
AGC122226  & 024638.9+274335&  9 & 625 & 8.6  &$-$13.13 & 7.56 &$-$13.17 &$-$13.06& $-$2.09 \\
AGES  & 030037.0+254707& 10 & 439 & 7.8  &$-$12.33 & 6.20 &$-$14.15 &$-$13.94& $-$3.06 \\
KKH18 & 030305.9+334140& 10 & 375 & 4.43 &$-$12.39 & 7.14 &$-$13.73 &$-$13.55& $-$3.16 \\
U2773 & 033207.1+474737&  9 & 397 & 5.5  &$-$16.12 & 8.27 &$-$11.83 &$-$11.30& $-$0.72 \\
U2905 & 035700.6+163128& 10 & 344 & 5.8  &$-$14.41 & 7.26 &$-$13.39 &$-$13.10& $-$2.47 \\
U3303 & 052459.5+043018&  8 & 446 & 7.2  &$-$16.03 & 8.55 &$-$11.95 &$-$11.80& $-$0.98 \\
KK49  & 054141.5+064054&  9 & 376 & 5.2  &$-$14.94 & 7.75 &$-$12.13 &$-$11.60& $-$1.07 \\
Orion & 054502.0+050406&  8 & 276 & 6.4  &$-$17.04 & 8.87 &$-$12.40 &$-$11.66& $-$0.95 \\
A0554 & 055736.7+072931& 10 & 340 & 5.5  &$-$12.85 & 7.30 &$-$13.92 &$-$13.37& $-$2.79 \\
HIZOA & 063009 :+082237& 10 & 259 & 3.6  &$- $9.7: & 7.17 &$-$14.0: &$-$13.4:& $-$3.2: \\
U3476 & 063029.2+331807& 10 & 477 & 7.0  &$-$14.27 & 8.15 &$-$12.74 &$-$12.52 &$-$1.73 \\
U3600 & 065540.0+390542& 10 & 435 & 7.3  &$-$13.53 & 7.84 &$-$13.52 &$-$13.44 &$-$2.61 \\
U3698 & 070918.8+442248& 10 & 464 & 7.2  &$-$14.30 & 7.78 &$-$13.03 &$-$12.94 &$-$2.13 \\
U3755 & 071351.8+103119& 10 & 190 & 6.98 &$-$15.53 & 7.89 &$-$12.64 &$-$12.56& $-$1.77 \\
DDO47 & 074155.0+164802&  8 & 160 & 7.98 &$-$16.04 & 9.06 &$-$12.57 &$-$12.54& $-$1.64 \\
KK65  & 074231.2+163340& 10 & 168 & 7.62 &$-$14.29 & 7.55 &$-$12.69 &$-$12.66& $-$1.79 \\
KK69  & 085250.7+334752& 10 & 419 & 7.7  &$-$11.96 & 7.59 &$-$15.2: &$-$15.2: &$-$4.3: \\
KK70  & 085522.0+333333& -3 &  -  & 7.7  &$-$11.86 & 6.8: &$-$15.3: &$-$15.2: &$-$4.4: \\
N2787 & 091918.6+691212&  1 & 838 & 7.48 &$-$18.50 & 8.05 &$-$12.54 &$-$12.33& $-$1.49 \\
N4600 & 124023.0+030704&  1 & 648 & 7.35 &$-$15.72 & 7.3: &$-$12.86 &$-$12.83& $-$2.00 \\
\hline
\end{tabular}
\end{table}

\clearpage
\hoffset=0cm
\begin{table}
\caption{External comparison of the measured $H\alpha$ fluxes}
\begin{tabular}{lccc} \hline
 Galaxy    & &                  $\log(F)$  &             \\
\hline
	&  This &   Zitrin \&   &  Kennicutt  \\
	 & paper & Brosch 2008 & et al. 2008       \\
\hline
A112521  &    $-$14.81   &     $-$14.36    &       $-$         \\
KK13     &    $-$13.77   &     $-$13.41    &       $-$         \\
KK14     &    $-$14.24   &     $-$14.53    &       $-$         \\
KK15     &    $-$14.52   &     $-$14.48    &       $-$         \\
IC1727   &    $-$11.94   &     $-$12.27    &   $-$11.96$\pm$.06  \\
N672     &    $-$11.56   &     $-$12.00    &   $-$11.49$\pm$.06   \\
U1281    &    $-$12.46   &     $-$12.65    &   $-$12.45$\pm$.07   \\
KK16     &    $-$13.97   &     $-$14.76    &      $-$         \\
KK17     &    $-$13.96   &     $-$14.03    &       $-$         \\
N784     &    $-$11.95   &     $-$11.66    &   $-$11.78$\pm$.04   \\
N855     &    $-$12.14   &     $-$12.29    &   $-$12.23$\pm$.04   \\
Maffei2  &    $-$11.99   &     $- $        &   $-$11.95$\pm$.06   \\
UA86     &    $-$11.94   &     $- $        &   $-$12.01$\pm$.07   \\
N1569    &    $-$10.59   &     $- $        &   $-$10.62$\pm$.01   \\
UA92     &    $-$12.30   &     $- $        &   $-$12.52$\pm$.03   \\
N1560    &    $-$11.60   &     $- $        &   $-$11.54$\pm$.05   \\
UA105    &    $-$11.58   &     $- $        &   $-$11.60$\pm$.03   \\
DDO47    &    $-$12.57   &     $- $        &   $-$12.51$\pm$.06   \\
\hline
\end{tabular}
\end{table}

\begin{table}
\caption{ Median parameters for different morphological samples }
\begin{tabular}{lcccc} \hline
 Type    &   $(T)$    &     $N$   &        $P$  &        $F$  \\
\hline
E,S0,dSph & ($<$1)     &    20   &     $-$1.7:  &     $-$0.4:   \\
         &          &         &           &            \\
Sa--Sbc    & (1--4)    &    26   &     $-$0.95  &     $-$0.55  \\
         &          &         &           &            \\
Sc--Sm    & (5--8)    &   108   &     $-$0.15 &     $-$0.10   \\
         &          &         &           &            \\
Irr,BCD   & (9--10)   &   240   &     $-$0.15 &     $+$ 0.25  \\
\hline
\end{tabular}
\end{table}

\begin{table}
\caption{Total SFR density in the Local universe
            ( z = 0, $H_0$ = 72 km s$^{-1}$ Mpc$^{-1}$, extinction corrected)   }
\begin{tabular}{cll} \hline
    $\log(\rho_{SFR})$   &  Reference                  &  Note           \\
    $M_{\odot}$/yr/Mpc$^3$ &                             &                 \\
\hline
    $-$1.95 $\pm$0.04   &  Gallego et al. 1995        &  emission line galaxies   \\
                   &                             &                 \\
    $-$1.73   $\pm$0.07   &  Tresse \& Maddox 1998       &  I--band survey  \\
                   &                             &                 \\
    $-$1.64   $\pm$0.02   &  Perez-Gonzalez et al. 2003 &  optically--selected  \\
                   &                             &                 \\
    $-$1.66   $\pm$0.08   &  Brinchmann et al. 2004     &  SDSS--based     \\
                     &                             &                 \\
    $-$1.81   $\pm$0.03   &  Hanish et al. 2006         &  HI-- selected   \\
                          &                             &                 \\
    $-$1.75   $\pm$0.03   &  Salim et al. 2007          &  UV--based       \\
                        &                             &                 \\
    $-$1.72   $\pm$0.08   &  James et al. 2008          &  $H\alpha$ Local universe\\
                     &                             &                 \\
    $-$1.72   $\pm$0.06   &  This paper               &  $H\alpha$ Local Volume \\
                          &                             &                 \\
 \hline
\end{tabular}
\end{table}

\begin{table}
\caption{Some cosmic density parameters }
\begin{tabular}{lcl} \hline
 Parameter       & Quantity                     &    Reference    \\
\hline
 $\rho_c$          & 1.43$\cdot 10^{11}     M_{\odot}$/Mpc$^3$    &    Spergel et al. 2007    \\
               &                               &                           \\
 $j_K(L)$         & 3.8$\cdot 10^8       L_{\odot}$/Mpc$^3$    &    Cole et al. 2001,  \\
               &                               &    Bell et al. 2003       \\
 $\rho(HI)$        & 0.44$\cdot 10^8      M_{\odot}$/Mpc$^3$    &    Zwaan et al. 2003      \\
               &                               &                           \\
 $\rho(SFR)$       & 0.019          $M_{\odot}$/yr Mpc$^3$ &    This paper          \\
               &                               &                           \\
 $\rho(P)$         &$-$0.17                          &    This paper          \\
               &                               &                           \\
 $\rho(F)$         &$-$0.50                          &    This paper          \\
               &                               &                           \\
\hline
\end{tabular}
\end{table}
\end{document}